# Angle-Resolved Cathodoluminescence Polarimetry of Hybrid Perovskites

Bibek S. Dhami[1†], Vasudevan Iyer[2†], Aniket Pant[1†], Ravi P. N. Tripathi[1],
Benjamin J. Lawrie[2,3] and Kannatassen Appavoo[1*]

[1]Department of Physics, University of Alabama at Birmingham, AL 35205
[2]Center for Nanophase Materials Sciences, Oak Ridge National Laboratory, TN, 37831, USA
[3]Materials Science and Technology Division, Oak Ridge National Laboratory, TN, 37831, USA

## Abstract

Coupling between light and matter strongly depends on the polarization of the electromagnetic field and the nature of the excitations in the material. As hybrid perovskites emerge as a promising class of materials for light-based technologies like LEDs, lasers, and photodetectors, understanding the microscopic details of how photons couple to matter is critical. While most optical studies have focused on the spectral content and quantum efficiency of emitted photons in various hybrid perovskite thin-film and nanoscale structures, few studies have explored other properties of the emitted photons such as polarization and emission angle. Here, we use angle-resolved cathodoluminescence microscopy to access the full polarization state of photons emitted from large-grain hybrid perovskite films with spatial resolution well below the optical diffraction limit. Mapping the Stokes parameters as a function of the emission angle in a thin film, we reveal the effect of a grain boundary on the degree of polarization and angle at which the photons are emitted. This exploration of angle- and polarization-resolved emission near grain boundaries provides an improved understanding of the emission properties of hybrid perovskites in thin film geometries — a necessary investigation for subsequent engineering of subwavelength nanophotonic structures using the hybrid perovskite class of materials.

## Keywords

[†]Authors contributed equally
[*]Address correspondence to appavoo@uab.edu

# Introduction

Hybrid organic-inorganic perovskites, referred to herein as hybrid perovskites, have emerged as a promising class of materials for various solution-processed thin-film technologies such as solar-to-energy conversion[1-5], sensing and telecommunications[6-8]. With absorption and emission signatures that can be easily tuned from the visible to the near-IR by composition or doping[8-10], this class of materials provides a playground for fundamental studies of a wide range of light-matter interactions under equilibrium (solar-like) and non-equilibrium (laser excitation) conditions. Although perovskite crystals have been studied for decades due to the exotic physics that arise from strong electron-correlation effects[11, 12], only recently was the organic-inorganic system stable enough to be considered for technological applications[11, 12]. This improvement in crystallinity and the resilience of hybrid perovskites to adverse stimuli such as moisture and humidity has resulted in groundbreaking power-conversion efficiencies exceeding 25% in single-junction photovoltaic devices[13, 14]. Benefiting from recent progress in photovoltaic device engineering, other device applications[2, 15-18] have been realized such as low-threshold nanolasers[8, 19], radiation detectors[7] and color-filters[6].

As hybrid perovskites find more applications in photonic technologies with critical dimensions at the nanoscale[20-23], it is clear that understanding the role of microstructures[24-26], grain boundaries[27-32], grain junctions[33] and interfacial crystal heterogeneities[34-36] is critical for enhancing device efficiency. In this regard, electron-beam spectroscopy techniques have been instrumental in characterizing hybrid perovskites below the diffraction limit[37-41]. For example, transmission electron microscopy (TEM) coupled with electron energy-loss spectroscopy (EELS) or energy-dispersive X-ray spectroscopy (EDX) has been employed to provide insights into the crystallization process, grain orientations with respect to the substrates, and ion migration towards boundaries[39]. Furthermore, electron backscatter diffraction (EBSD) measurements can probe crystallographic information and provide insights into the distribution of actual grain sizes created by various fabrication techniques. EBSD has also been used to probe the influence of grain-orientation heterogeneity (sub-grain boundaries and intra-grain misorientation) on local strain[34] and non-radiative recombination pathways[35] that ultimately determine the emission efficiency. Recently, by probing the crystallinity of hybrid perovskites at grain boundaries, EBSD has revealed the presence of amorphous grain boundaries that give rise to brighter emission as a consequence of longer carrier lifetimes[28]. Cathodoluminescence (CL) microscopy has been used extensively to probe structure-function relationships of hybrid perovskites at the nanoscale[37, 42-45] in order to describe how fine tuning of materials properties in halide perovskites can increase device efficiency and improve long-term stability. In nanophotonics, for example, CL microscopy has been used to map nanoscale electromagnetic modes in plasmonic and all-dielectric metamaterial systems[37, 46-48]. Furthermore, CL microscopy has been used to probe other physical processes such as Purcell enhancement using plasmonics[49-53], dispersion of quasiparticles such as surface plasmon polaritons[54, 55] and collective Bloch modes in photonic crystals[56-59].

In this work, we adapt a recently developed technique in cathodoluminescence spectroscopy to access the full polarization state of emission from a large-grain hybrid perovskite thin film. We use a focused energetic electron beam as a broadband excitation source to allow us interrogate nanoscale regions of interest. By tuning the electron-beam energy (across $1-30\ keV$), we can explore excited state energetics with nanoscale depth resolution in addition to the nanoscale lateral resolution provide by the converged electron beam. Thanks to this high degree of spatial resolution, typically $< 10\ nm$, we efficiently excite and probe emission with a relatively low background signal, allowing us to pinpoint the effects of grain morphology and boundaries on emission properties. Similar to dark-field techniques, the emission generated from this excited region dominates the Fourier image, allowing us to determine the full angle-resolved polarization state of light emitted from these films. Encoded in the polarization of the emitted photons are details of the perovskite photophysics, including the local orientation of emission centers and symmetry-breaking structural changes in the film under investigation. To completely describe the polarization state of the light emission from our hybrid perovskite, we use the Mueller matrix formalism to determine the Stokes parameters of the emitted photons at the sample plane. We find that the intensity of the unpolarized light is strongly modulated by the presence of grain boundaries, though the angular emission profile is largely unchanged by grain boundaries. Unexpectedly, we observe a strong angular dependence for polarized CL at the grain boundary, highlighting the complex emission mechanisms that occur at the nanoscale.

## Cathodoluminescence Microscopy

To access full polarization information in cathodoluminescence spectroscopy, we employ a technique pioneered by the Polman and Koenderink groups, originally based on polarization analysis in optical microscopes[60]. An electron beam with energy between $1 - 30 \, keV$ excites the sample in an environmental scanning electron microscope. An aluminum parabolic mirror collects the CL emission from the sample and projects the emission on a two-dimensional $1024 \times 256$-pixel charge-coupled device (CCD) array that measures the intensity profile of the incoming emission beam. The distribution of the wavevectors in CL emission can be retrieved from the CCD image as each emission angle corresponds uniquely to a pixel on the CCD. As shown in Figure 1, the emitted photons can be analyzed using different modalities including hyperspectral, polarized and angle-resolved imaging with a spatial resolution that is well below the diffraction limit. A rotating-plate polarimeter is included in the optical path that consists of a quarter-wave plate and a linear polarizer (Figure 1). The polarization of the light can be measured by varying the orientation of these optical elements. Importantly, the angle-dependent four Stokes parameters of the light can also be calculated, providing the most general representation of the polarization state of the cathodoluminescence[60, 61]:

$$S_0 = |\hat{\epsilon}_1 \cdot \vec{E}|^2 + |\hat{\epsilon}_2 \cdot \vec{E}|^2 = I(0^o, 0) + I(90^o, 0)$$
$$S_1 = |\hat{\epsilon}_1 \cdot \vec{E}|^2 + |\hat{\epsilon}_2 \cdot \vec{E}|^2 = I(0^o, 0) - I(90^o, 0)$$
$$S_2 = 2 \, \mathfrak{Re}\left[ (\hat{\epsilon}_1 \cdot \vec{E})^*(\hat{\epsilon}_2 \cdot \vec{E}) \right] = I(45^o, 0) - I(135^o, 0)$$
$$S_3 = 2 \, \mathfrak{Im}\left[ (\hat{\epsilon}_1 \cdot \vec{E})^*(\hat{\epsilon}_2 \cdot \vec{E}) \right] = I\left(45^o, \frac{\pi}{2}\right) - I\left(135^o, \frac{\pi}{2}\right)$$

written in terms of general polarization basis sets that are orthogonal to each other. Here we choose $\hat{\epsilon}_1 = \hat{\epsilon}_\vartheta$ and $\hat{\epsilon}_2 = \hat{\epsilon}_\varphi$ for the detected emission from the source-sample frame. Thus, following detection of these raw-polarized images on our CCD (i.e., detector plane), the images are then projected onto the zenithal (also referred as radial, $\vartheta$) and azimuthal ($\varphi$) space. To transform these to Stokes parameters in the sample plane, we use the appropriate Mueller matrix of the light collection system that accounts for the effect of the parabolic mirror on the polarization, where each element in this matrix is a function of the emission angle and wavelength. Included in this analysis are the effects of $s$- and $p$-polarized light based on the Fresnel coefficients of the mirror at the central frequency of the collection bandwidth.

## Cathodoluminescence Polarimetry on Hybrid Perovskites

To understand how the morphology of a grain and its boundaries modify the intensity and polarization properties of the emission process, we investigate a large-grain hybrid perovskite with maximum emission near its band edge ($\lambda \sim 760 \, nm$). Fabrication and characterization of the hybrid perovskite, here methylammonium lead iodide ($CH_3NH_3PbI_3 = MAPbI_3$), are described in the $Methods$ section and in references[32, 62]. Initially, we use spatially resolved CL spectroscopy by combining scanning electron microscopy with the detection of CL that is emitted from the sample (Figure 1c,d). We perform the CL mapping and polarimetry using a $5\ keV$, $110\ pA$ electron beam. CL mapping was performed by collecting one spectrum for each pixel with an acquisition time of $100\ ms$ to form a $256 \times 256$-pixel image of a hybrid perovskite domain (Figure 1c). A line cut through the SEM/CL map along a grain boundary shows the change in emission spectrum as a function of location of the electron-beam excitation on the hybrid perovskite (Figure 1d,e).

Because spatially resolved maps of the CL polarization state require increased electron beam doses, a simple understanding of electron-beam-induced damage in $MAPbI_3$ is critical. We find that for large beam energy ($> 10\ keV$) or current ($> 2\ nA$), spectral changes over time were visible within the first 10 seconds of CL collection. CL peak broadening accompanied by a decrease in peak intensity most likely resulted from nonradiative decay pathways at defects created by the electron beam. Furthermore, when nanoscale areas of the hybrid perovskite are exposed to the $10\ keV$ or $20\ keV$ electron beam for 10s of seconds, two new CL peaks emerge at higher photon energy. A CL peak at $\sim 660\ nm$ ($1.88\ eV$) is attributed to intermediate phases that are formed during the perovskite decomposition and caused by electron-beam-induced heating. Hybrid perovskites are poor electrical and thermal conductors and such a focused beam ($\sim 5\ nm$) could conceivably create local heating effects. The CL peak at the highest photon energy of $525\ nm$ ($2.36\ eV$), corresponds to the presence of $PbI_2$. The $PbI_2$ CL is stable over extended periods under exposure to the electron beam and is not reversible, unlike changes caused by an intense laser source[63]. These observations are well aligned with the extensive work by Xiao et al.[64], where the authors conducted in situ studies of high-energy electron beam interaction with hybrid perovskites. Two damage mechanisms were proposed, including nanoscale local heating and ion displacement via the knock-on or Frenkel defect mechanism. Importantly for our studies, it was found that an irradiation power of $\sim 5 \times 10^{10}\ W/m^2$ would minimize damage to the hybrid perovskite. Therefore, for the remainder of this work, we excite the hybrid perovskite with electron beam having power that lies in the $0.5 - 1 \times 10^{10}\ W/m^2$ range.

**Results and Discussions**

Figure 2 illustrates the normalized angle-dependent, four Stokes parameters at the sample plane ($S_0, S_{1N}, S_{2N}, S_{3N}$) that have been calculated from the six polarization measurements described above. It provides a comparison between the two extreme cases in our film, i.e., emission collected from a grain center (left panels) and emission collected at the grain boundary (right panels), with the topmost panels corresponding to the total intensity distribution, $S_0$. In order to observe features in the polarization at all angles, all the panels have been normalized with respect to $S_0$. In both $S_0$ cases, the hybrid perovskite semiconductor film emits in a Lambertian distribution, a direct consequence of Snell's Law[65]. This emission distribution with its characteristic cosine dependence on the zenithal angle is expected for semiconductors and dielectrics because they radiate incoherently in the material via spontaneous emission[66]. Here, the effects of the grain boundary on the emission polarization are readily seen in the parameters $S1N$ and $S2N$ that are highly dependent on the emission angle, while for the grain center, all emission pattern plots are fairly monotonic in nature, with little angular dependence.

Having retrieved the Stokes parameters for our sample, we are now in a position to compute the electric field components. To understand how maps of far-field emission are derived from local radiating sources, we reconstruct the spherical field vector amplitudes $|\hat{\epsilon}_\vartheta \cdot \vec{E}|$ and $|\hat{\epsilon}_\varphi \cdot \vec{E}|$ that correspond to the intrinsic parallel ($p$-) and perpendicular ($s$-) polarization basis set relevant to our system, with respect to the plane of incidence as defined by the propagation vector ($\vec{\kappa}$). Figure 3 shows that, at the grain center, the $|\vec{E}_\vartheta|$ and $|\vec{E}_\varphi|$ distributions are strong and azimuthally symmetric. This feature is expected since the filtered light ($\lambda = 750 \pm 25\ nm$) originates from the incoherent spontaneous emission process within the bulk hybrid perovskite. However, at the grain boundary where similar results are expected, we see that the electric-field amplitude is angle dependent. These results highlight the complex nature of grain boundaries and their role in strongly modulating the emission direction. These novel results complement the well-studied changes in spectral content and intensity of the emitted photons that occur at grain boundaries and junctions[28, 35, 67]. Furthermore, the polarization can also be disentangled in Cartesian space, i.e., using vector amplitudes $|\vec{E}_x|$, $|\vec{E}_y|$ and $|\vec{E}_z|$. Plotted this way, we see that emission in the sample plane has a complex angular dependence, with a four-lobe pattern. For either selected region of interest (center or boundary), we see complimentary polarization in the $|\vec{E}_x|$ and $|\vec{E}_y|$ direction. Moreover, for the grain boundary case, this polarization pattern appears stronger than for the grain center, caused by the interfacial boundary that presumably leads to in-plane symmetry breaking. Taking a look at the $|\vec{E}_z|$ component, we see a characteristic doughnut shape, corresponding to a decrease in that component as $\vartheta$ approaches zero. In other words, since the electric field is necessarily transverse to the propagation vector $\vec{\kappa}$, $|\vec{E}_z|$ vanishes at near-normal angles. Furthermore, for the grain boundary case, we see an overall decrease in the magnitude of this component, accompanied by a narrower range of emission angle. Two potential factors for a decrease in emission are a reduction in local sources of emission at the boundary (i.e., increase in non-radiative decay channels), and emission from local dipoles that are deeper in the grain (but exposed thanks to the presence of the boundary). Both will result in greater scattering of the emission since the feature sizes of the interface are on the same order as the emission wavelength.

The degree of linear polarization (DOLP) and the degree of circular polarization (DOCP) can also be calculated from the measured Stokes parameters. These parameters are given by the ratio of polarization with respect to the total intensity and thus, DOLP is given by $\sqrt{(S_1^2 + S_2^2)}/S_0$ and DOCP is given by $S_3/S_0$. Figure 4 highlights the difference in DOLP and DOCP between the center and boundary of the grain. At the boundary, the symmetry breaking in the sample plane causes linear polarization to increase as $\vartheta$ approaches $90^0$. Moreover, for an in-plane angular range of $\varphi$ $(0 - 90^0)$, we see that there is also an increase in the DOLP, likely because the electron beam is not perfectly at the center of the grain boundary. We note that for $\vartheta > 60^0$, the DOLP appears higher for the grain boundary case. A similar effect is seen for the DOCP, whereby the grain boundary exhibits a greater degree of circular polarization (negative in this case). For the case of the grain center, the relatively flat morphology of the thin film results in a negligible DOCP as expected from the isotropic spontaneous emission process.

In figure 4, we average the azimuthal component of the emission ($\varphi$ ranging from $90^0 - 270^0$, $counterclockwise$) and plot the zenithal cross cuts for both the polarized ($S_0 \times DOP$) and unpolarized ($S_0 \times (1 - DOP)$) cases. The total angle-resolved CL is plotted in figure 5a, demonstrating that indeed emission from the hybrid perovskite is mainly unpolarized with a Lambertian-type profile; CL is reduced at the grain boundary either due to an increase in non-radiative decay channels or a decrease in local emission sources caused essentially by less hybrid perovskite material being present at the boundary. Furthermore, we see that polarized emission is lower than its unpolarized counterpart in both cases; it is about an order of magnitude less for the grain center and about three times less for the grain boundary. Since the collected emission is from band-to-band recombination at $\lambda \sim 760 \, nm$, the emission inside the bulk thin film is incoherent, unpolarized and isotropic. Thus, the slight polarized emission for the grain center data (red solid line) collected by our system is a result of the semiconductor-vacuum interface with large differences between the $s$- and $p$-Fresnel transmission coefficients. The angular distribution of this weak polarized emission is very different from the unpolarized emission and agrees with Fresnel calculations in some measure, which display an increase of $Tp$ vs. $Ts$ as the outgoing angle increases, i.e., as $\vartheta$ approaches $90^0$. This angular emission profile is rather complex as it is attributed both to changes of polarization at the interface, and to other factors such as changes in the total internal reflection caused by the film morphologies and restricting emission at certain angles. Furthermore, photon-recycling can also play a role in modifying the radiative properties of hybrid perovskites[20, 68], and could redistribute the angular emission profile within the spectral content studied. Because hybrid perovskites have strong bandedge CL and the collected emission was filtered, coherent transition radiation should play little role in the angular profile of weak polarized emission. A similar conclusion was reached for GaAs[60, 66]. An important observation from Figure 5 is that the weak polarized emission from the grain boundary is peaked both at $\vartheta = 45^0$ and unexpectedly at $\vartheta = 0^0$. We attribute emission for the latter to the vertical surfaces of the grain boundary (i.e., perpendicular to the substrate) that polarizes the emission in a similar fashion as the grain center. Consequently, the additional emission at the boundary results in this peculiar angular emission profile (Figure 5b, blue). A careful look at Figure 5a, which recorded the total emission count, shows that the *polarized* emission at the boundary is indeed higher than the emission intensity at the grain center.

**Conclusions**

Our data show that precise electron-beam excitation and polarimetry analysis of the CL captured provides quantitative information about the origin of emission in large-grain hybrid perovskite thin films. We use this novel technique to map the vectorial electromagnetic emission properties of this nanostructured thin film, and importantly separate polarized and unpolarized emission, which can also be used to determine different mechanisms that contribute to emission processes in this technologically relevant material. Thanks to the high resolution of the electron beam excitation, the wave-vector resolved polarization properties of locally excited emissive events can be extracted with a spatial resolution of $< 10\ nm$. Moreover, we also demonstrate how structural asymmetry translates into changes in linear and circular polarized emitted light. This demonstration of nanoscale characterization highlights the advantages of SEM-CL to locally study material anisotropy and optical activity near grain boundaries.

## Methods

**Fabrication of hybrid perovskite:** Hybrid perovskite thin films were prepared using hot-casting technique[62]. The perovskite precursor solution is made by mixing $0.2\ gram$ methyl ammonium iodide ($CH_3NH_3PbI_3 = MAPbI_3$, Sigma Aldrich, 98 %) and $0.578\ gram$ lead iodide ($PbI_2$, Sigma Aldrich, 99 %) in $1.0\ ml$ of anhydrous, N, N-Dimethyl formamide ($DMF$, Sigma Aldrich, 99 %). The mixture solution was then heated at $120^oC$. We used borosilicate microscopic glass slides as the substrate. The glass substrates are cleaned by isopropanol (Sigma Aldrich, 99 %) using ultrasonication. We heated these cleaned substrates to $300^oC$. We then spin-coated the preheated perovskite-precursor solution onto preheated substrate at $5000\ rpm$ for $5\ seconds$ and quickly transferred the substrate on the hotplate maintained at $300^oC$ for $2\ seconds$. Subsequently, we obtained thin film of hybrid perovskite with large grain features. This fabrication process was performed inside a $N_2$ filled glovebox.

**Cathodoluminescence microscopy:** CL microscopy was performed in a FEI Quattro environmental scanning electron microscope with a Delmic Sparc cathodoluminescence system. A beam energy of $5\ keV$ and beam current of $110\ pA$ were used. The sample chamber was maintained at a pressure of $2 \times 10^{-3}\ Pa$. CL spectra were acquired with a high numerical aperture (0.9) parabolic mirror and sent to an Andor Kymera 193i spectrometer equipped with a $150\ line/mm$ grating and an Andor Newton CCD camera. For the angle-resolved measurements, the sample was exposed for $500\ ms$ at each polarizer orientation, resulting in a total exposure of $3\ seconds$ per acquisition. The Fourier plane of the image is obtained by using an appropriate lens and switching to a mirror in the spectrometer. The exposure time for standard spectrum imaging was $100\ ms$ per pixel.

## Author Contributions

BSD and RPNT fabricated the hybrid perovskite samples. VI conducted the CL microscopy. All authors help analyze the data and contributed to writing the manuscript. KA and BJL designed and oversaw the project.

## Acknowledgment

Support for this project was provided by NASA EPSCoR RID (award number 80NSSC19M0051) and UAB startup funds. Cathodoluminescence microscopy was conducted at the Center for Nanophase Materials Sciences, which is a DOE Office of Science User Facility. BSD acknowledges financial support from the Alabama Graduate Research Scholars Program (GRSP) funded through the Alabama Commission for Higher Education and administered by the Alabama EPSCoR.

# Figures

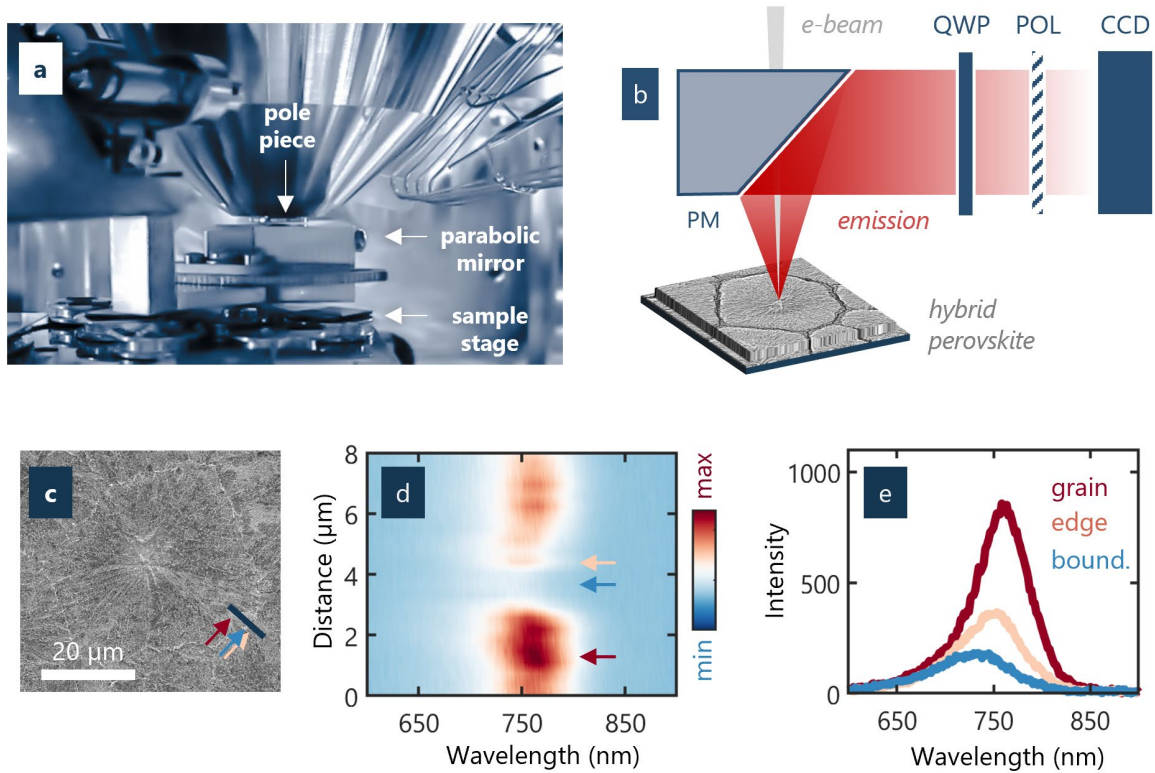

**Figure 1: Overview of cathodoluminescence (CL) polarimetry of hybrid perovskite. (a)** Photograph of scanning electron microscope (SEM) chamber with CL polarimetry capabilities. **(b)** Schematics of angle-resolved CL polarimetry of large-grain hybrid perovskite where emitted photons are directed to the detector using an off-axis paraboloid mirror, analyzed using rotating-plate optical polarimeter components placed in the beam path. PM: Parabolic mirror; QWP: Quarter wave plate; POL: Polarizer; CCD: charged-coupled device. **(c)** Scanning electron micrograph depicting one of the probed hybrid perovskite grains. Dark blue line represents a spatially selected region interest, here a grain boundary. **(d)** CL spectrum as a function of scanned distance. Pseudo-color plot of emission wavelength vs. spatial distance, with the suppressed axis representing intensity of the CL. **(e)** For the three arrows drawn in figure (c, d), typical spectra at the grain center (red), edge (pink) and boundary (blue) are plotted. Note the non-negligible emission at the grain boundary that systematically shows a blue shifted peak.

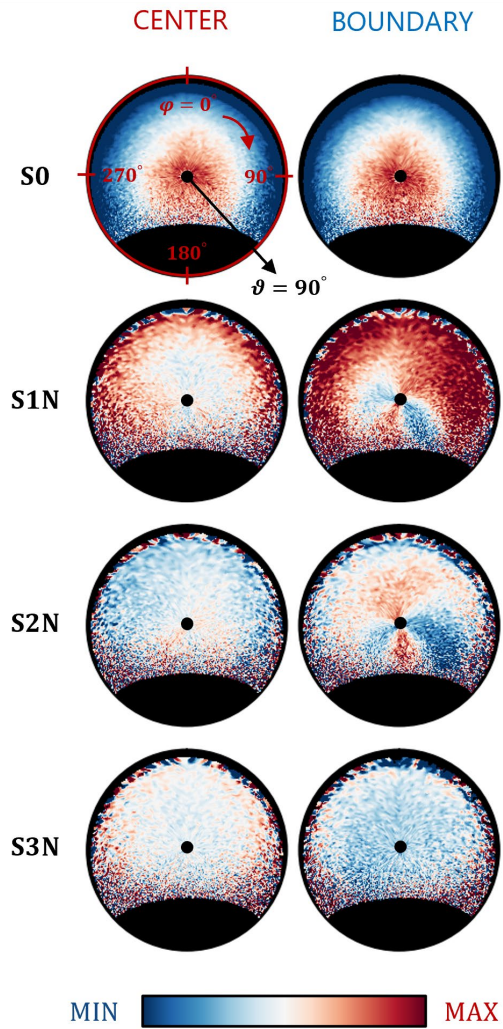

**Figure 2: Stokes parameters of CL in hybrid perovskite.** Angle-resolved Stokes parameters in the sample plane comparing CL from the grain center (left column) and CL from the grain boundary (right column) mapped onto the polar coordinates (azimuthal angle $\varphi$ and zenithal angle $\vartheta$). The $S1N, S2N$ and $S3N$ parameters have been normalized to their respective $S0$ to easily compared between the overall polarization distribution. For example, strong linear polarization dependence ($S1N$ and $S2N$) is visible for the grain boundary case.

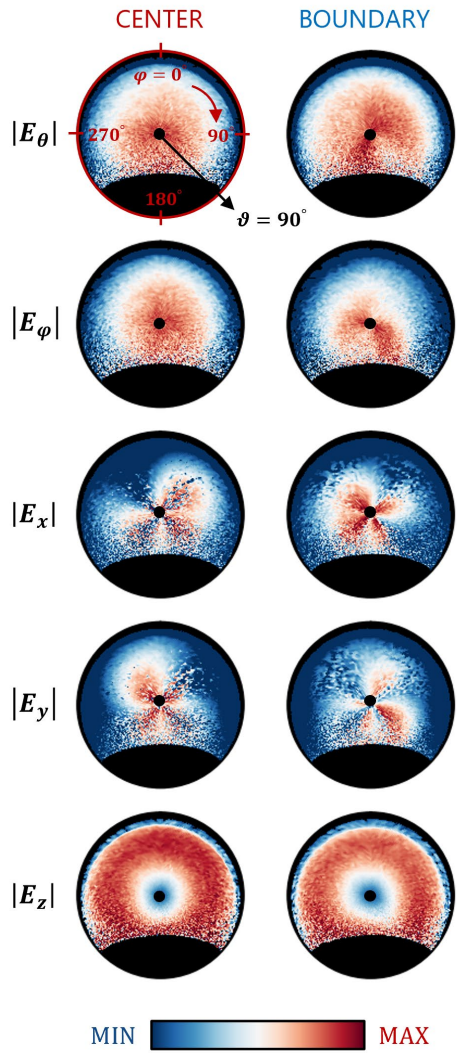

**Figure 3: Retrieved electric field amplitude distributions.** Angle-resolved spherical and Cartesian electric field amplitude distributions for CL from the grain center (left column) and from the grain boundary (right column). Note that the amplitudes $|\vec{E}_\varphi|$ and $|\vec{E}_\vartheta|$ corresponds to the $s$- and $p$- polarization basis that map the far-field generated by a localized emission event and are the components that highlight differences in the overall emission as a function of angle.

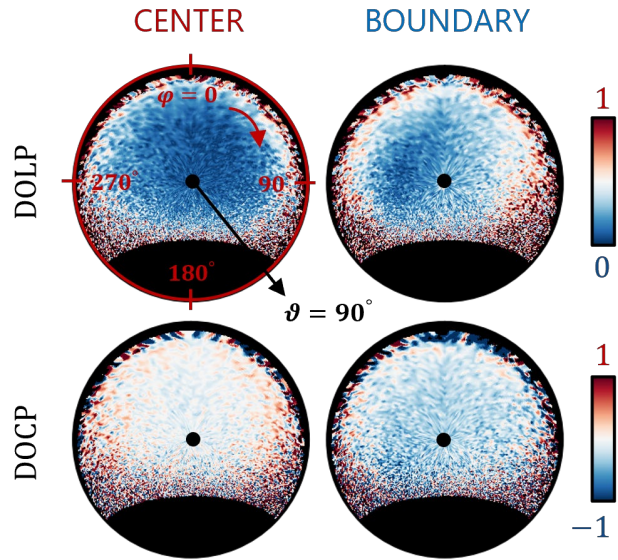

**Figure 4: Comparison of degree of polarization in CL.** Degree of linear polarization (DOLP) and degree of circular polarization (DOCP) between CL from grain center (left column) and grain boundary (right column) as a function of angle. For the boundary case, DOLP shows an asymmetrical emission profile while the DOCP plot shows a slightly negative polarization dependence (blue).

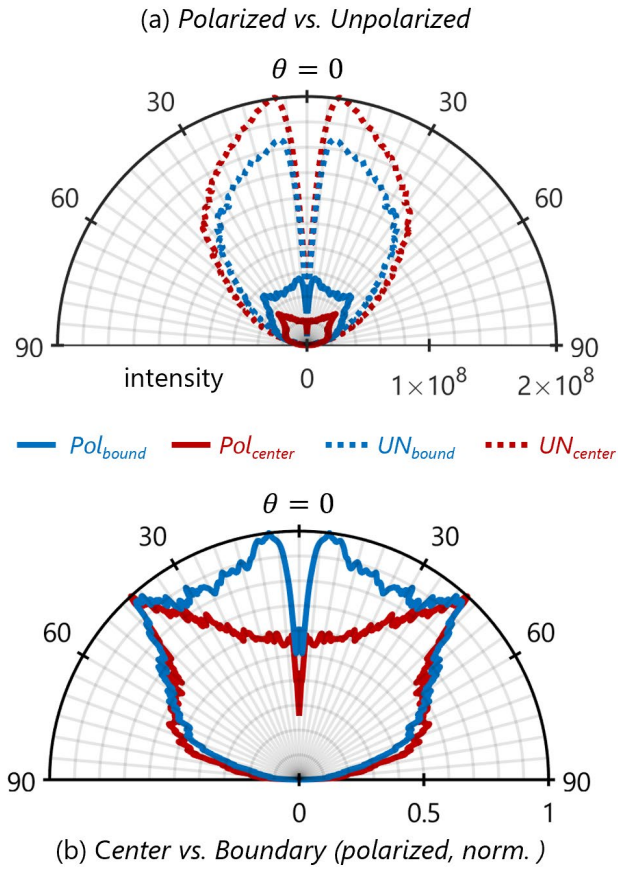

**Figure 5: Polarized and unpolarized zenithal emission from hybrid perovskite. (a)** Zenithal cross cuts comparing polarized (line) and unpolarized (dash) emission counts from the grain center (red) and grain boundary (blue) at $\lambda = 750\ nm$. **(b)** Normalized zenithal cross cuts to highlight effect of the boundary on the polarized emission, where we scale the angular distributions by the overall polarized emission intensity. The data has been obtained by averaging over the azimuthal range $\varphi = 90^o - 270^o, counterclockwise$ to improve signal-to-noise ratio.